# Proximity superconductivity in atom-by-atom crafted quantum dots

Lucas Schneider[1,*], Khai That Ton[1], Ioannis Ioannidis[2,3], Jannis Neuhaus-Steinmetz[1], Thore Posske[2,3], Roland Wiesendanger[1], and Jens Wiebe[1]

[1]Department of Physics, University of Hamburg, D-20355 Hamburg, Germany.

[2]I. Institute for Theoretical Physics, University of Hamburg, D-20355 Hamburg, Germany.

[3]Centre for Ultrafast Imaging, Luruper Chaussee 149, D-22761 Hamburg, Germany.

[*]E-mail: lucas.schneider@physnet.uni-hamburg.de

**Abstract**
**Gapless materials in electronic contact with superconductors acquire proximity-induced superconductivity in a region near the interface[1,2]. Numerous proposals build on this addition of electron pairing to originally non-superconducting systems like ferromagnets and predict intriguing quantum phases of matter, including topological-[3–7], odd-frequency-[8], or nodal-point[9] superconductivity. However, atomic-scale experimental investigations of the microscopic mechanisms leading to proximity-induced Cooper pairing in surface or interface states are missing. Here, we investigate the most miniature example of the proximity effect on only a single quantum level of a surface state confined in a quantum corral[10,11] on a superconducting substrate, built atom-by-atom by a scanning tunneling microscope. Whenever an eigenmode of the corral is pitched close to the Fermi energy by adjusting the corral's size, a pair of particle-hole symmetric states enters the superconductor's gap. We identify the in-gap states as scattering resonances theoretically predicted 50 years ago by Machida and Shibata[12], which had so far eluded detection. We further show that the observed anticrossings of the in-gap states indicate proximity-induced pairing in the quantum corral's eigenmodes. Our results have direct consequences on the interpretation of in-gap states in unconventional or topological superconductors, corroborate concepts to induce superconductivity into a single quantum level and further pave the way towards superconducting artificial lattices.**

**Main**
Combining the individual properties of different quantum materials in hybrid structures offers a seemingly inexhaustible variety of exotic phases of matter[13], including strongly correlated electron systems[14], topologically non-trivial spin textures[15], quantum anomalous Hall effects[16], or unconventional superconductivity[4,6–8]. Particularly interesting states are formed when superconductivity (SC) is induced into intrinsically non-superconducting materials by the proximity effect[1,2], giving rise to topological-[3–7], spin-triplet-[8], nodal-point-[9] or Fulde–Ferrell–Larkin–Ovchinnikov-SC[17]. A good understanding and control of the proximity effect in metallic nanostructures is crucial for the development of such novel heterostructures. Pairing in the normal metal is induced via Andreev reflection processes at the interface with the superconductor. If the transparency of the interface between a normal metal in the clean limit and the superconductor is high, SC is induced over a length scale which can exceed dozens of nanometers[18]. However, for many of the exciting heterostructures, SC has to be induced through interface states or into surface states[6,19–22]. These are typically well decoupled from the bulk bands and thus it is unclear *a priori* whether they will acquire sufficient pairing if their distance to the superconductor is larger than a few nanometers[18,22]. To study this effect in detail, we downscale the problem as much as possible by investigating only a single resonance mode of a surface state. This is achieved by laterally confining the surface state in a quantum corral, forming a particular quantum dot (QD). These can naturally occur in nanoscopic islands[23,24] or, in a more tunable platform, in artificially designed adsorbate arrays[10,25,26] where the QD walls are built atom-by-atom using the tip of a scanning tunneling microscope as a tool. Although the surface states are typically well decoupled from metallic bulk states in the direction perpendicular to the surface plane, scattering at step edges or the adsorbates is known to introduce a measurable coupling to the bulk electronic states leading to a lifetime broadening of the QD's eigenmodes $\Gamma$ on the order of several meV[27,28]. Notably, in contrast to the more widely studied semiconductor or molecular QDs[29], the electron density screening the metallic QDs investigated here is by orders of magnitude larger, which leads to largely suppressed electron-electron interactions, i.e., the QD charging energy $U$ is negligible, and thereby, the QD can be described by spin-degenerate single-particle eigenmodes. Recently, coupled arrays of such QDs with tunable interactions between adjacent sites have evolved as an exciting platform for the simulation of quantum materials[30,31]. However, while there has been progress in choosing different material templates for incorporating more complex phenomena like, e.g., Rashba spin-orbit coupling into these QDs[32], pathways for inducing SC into their individual eigenmodes have not been studied so far. To this end, artificial lattices including SC pairing terms are among the most promising platforms for the realization and control of topological superconductivity and in particular of Majorana zero modes[33,34].

Here, we study artificial QDs defined by a cage of Ag atoms on thin Ag(111) islands (see Figs. 1a and b, Methods, and Supplementary Note 1) grown on superconducting Nb(110) using scanning tunneling microscopy (STM) and -spectroscopy (STS). We employ superconducting Nb tips for enhanced energy resolution in the STS experiments. The use of Nb tips leads to a shift of spectral features to higher energies by the value of the tip's superconducting gap $\Delta_\mathrm{t}$, i.e., states at the sample's Fermi energy $E_\mathrm{F}$ are found at bias voltages of $e \cdot V = \pm \Delta_\mathrm{t}$. The proximity to Nb(110) opens a superconducting gap of $2\Delta_\mathrm{s}$ in the bulk states of Ag(111) for island thicknesses well below $d_\mathrm{Ag}$ = 100 nm[18,35]. We measure a value of $\Delta_\mathrm{s}$ = 1.35 meV (Supplementary Note 2), which is similar to the gap of elemental Nb, $\Delta_\mathrm{Nb}$ = 1.50 meV[3,36], indicating a high interface quality between Nb and Ag. The outline of the experiment is shown in Fig. 1b: the scattered Ag(111) surface state electrons visible as wavy patterns at the surface of Ag islands (Fig. 1a) are confined within a couple of lattice constants in the direction perpendicular to the surface[37] but still have a finite coupling $V \propto \sqrt{\Gamma}$ to the superconducting Ag bulk electrons[28,38]. We further confine these electrons laterally within QDs built of walls of Ag atoms resulting in spin-degenerate eigenmodes of energies $E_\mathrm{r}$ which can be pitched to the Fermi energy $E_\mathrm{F}$ by adjusting the width $L_\mathrm{x}$ of the QD. We then investigate the proximity effect of the bulk electrons onto these QD eigenmodes.

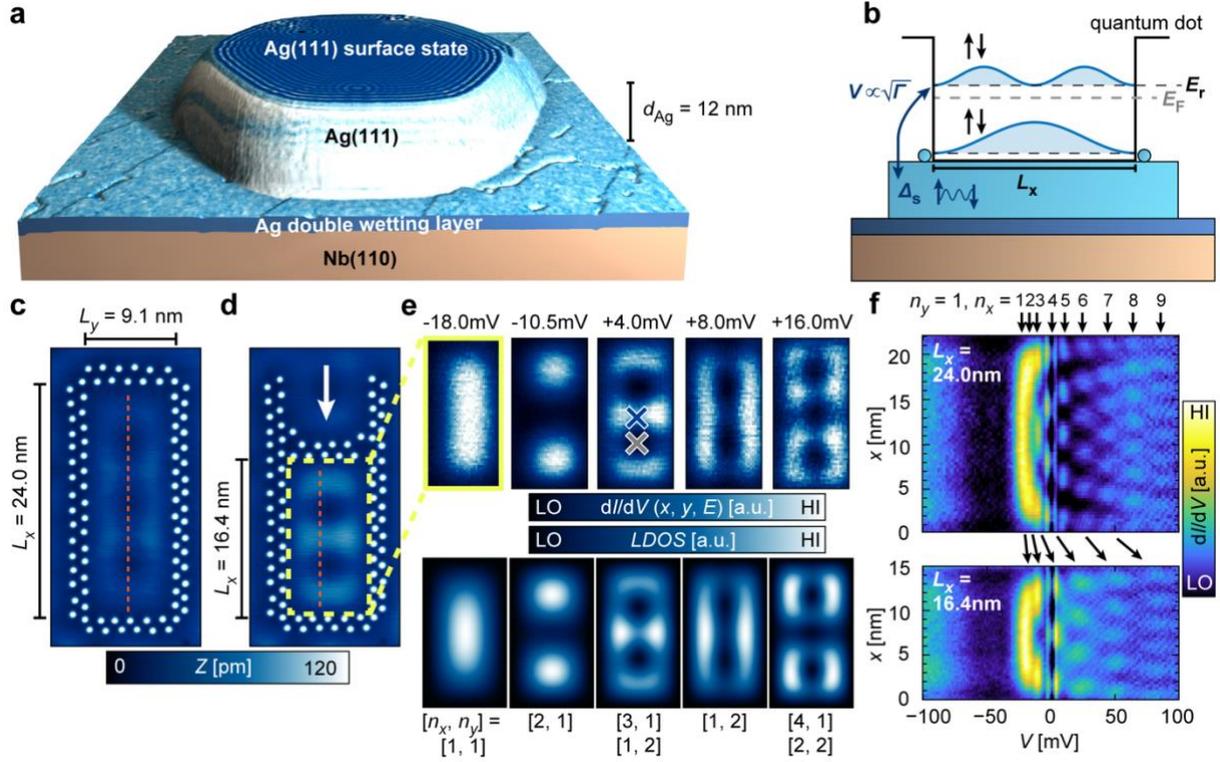

**Fig. 1 | Atom-by-atom built quantum dots coupled to the superconducting substrate. a**, 3D rendering of the constant-current STM topography of a Ag island with a thickness of 12 nm. The simultaneously measured d$I$/d$V$ signal is used as the 3D model's texture, showing quasiparticle interference patterns of the surface state electrons on the Ag(111) surface termination. The island grows on top of a pseudomorphic Ag double layer on Nb(110) (sketched profile, see Supplementary Note 1 for details). **b**, Sketch of the experimental setup with the QD having walls built of single atoms, laterally confining the surface state electrons such that spin-degenerate QD eigenmodes of energies $E_r$ are formed. The QD eigenmodes couple to the gapped superconducting substrate ($\Delta_s$) with a strength $V \propto \sqrt{\Gamma}$. $E_r$ can be pitched with respect to the Fermi energy $E_F$ by adjusting the width $L_x$ of the QD. **c**, Constant-current STM image of a rectangular QD with side-lengths $L_x$ and $L_y$ consisting of 44 non-magnetic Ag atoms. $L_x$ and $L_y$ are defined as the distance between the Ag atoms in the inner ring. The QD is surrounded by another 44 Ag atoms preventing surface state modes located outside the QD from leaking into the structure. **d**, Constant-current STM image of the same structure with one of the QD walls moved to the lower end as indicated by the white arrow. **e**, Upper panels: Constant-height d$I$/d$V$ maps at bias voltages indicated in the respective panels measured in the interior of the QD in panel d (area marked by the yellow dashed lines). All panels are 15 x 7.5 nm$^2$ in size. Lower panels: Simulation of a rectangular box with infinitely high potential walls and dimensions $L_x$ = 16.4 nm, $L_y$ = 9.1 nm assuming a parabolic dispersion of the quasiparticles in 2D with $m_{eff}$ = 0.51$m_0$ and $E_0$ = -30.3meV (see Supplementary Note 4 for details). The quantum numbers [$n_x$, $n_y$] of the dominant eigenmodes at the energies of the experimental maps are given below each simulated map. Note, that some maps are a mixture of two eigenmodes because of energy level broadening. **f**, d$I$/d$V$ line profiles along the dashed orange vertical lines of the two QDs marked in panels c and d obtained in constant height mode. QD eigenmodes with $n_y$ = 1 and $n_x$ as indicated by the arrows above the panel are observed. Their respective energy is shifted when the length $L_x$ is altered as illustrated by the black arrows. Parameters: $V$ = 50 mV, $I$ = 1 nA, $V_{mod}$ = 5 mV for panel a; $V$ = 5 mV, $I$ = 1 nA for panels c and d; $V_{stab}$ = -100 mV, $I_{stab}$ = 2 nA, $V_{mod}$ = 2 mV for e and f.

**Design and pitching of quantum dot states**

As shown by Limot *et al.*[39], individual Ag atoms can be reproducibly gathered by approaching the STM tip close to the Ag(111) surface (see Supplementary Note 3 for details). These can be arranged to form rectangular artificial QDs of tunable sizes (Figs. 1c, d) using lateral atom manipulation techniques (see Methods). Since the Ag walls of the QDs have a finite transparency for the surface state electrons, a second wall of Ag atoms is constructed around the central QD wall in order to screen the interior from surface state modes located outside of the structure. The spatial structure of the QD's eigenmodes can be mapped by measuring the differential conductance d$I$/d$V$($x$, $y$, $E$) at a particular bias voltage $eV = E$. The resulting patterns (Fig. 1e, upper panels) closely resemble the eigenmodes of a two-dimensional rectangular box potential with infinite walls having a well-defined number of antinodes in $x$ and $y$ direction [$n_x$, $n_y$] (Fig. 1e, lower panels, see Supplementary Note 4 for details). In the following, the width $L_y$ of the QD is kept fixed while the length $L_x$ is tuned by moving the upper Ag wall laterally (see Fig. 1d). This leads to a change in the confinement conditions such that the eigenenergies of the QD states are shifted. Experimentally, this can be verified by measuring d$I$/d$V$ line profiles along lines close to the central axis of a given QD (Fig. 1f, upper panel): the eigenmodes with $n_y$ = 1 and $n_x$ = 1, 2, 3, … can be identified and are marked by black arrows. When the QD length $L_x$ is changed from 24.0 nm to 16.4 nm (lower panel), a shift of the individual states to higher energies can be observed (black arrows)[26]. Note that it can be already seen by comparison of the top and bottom panels of Fig. 1f that, decreasing the length $L_x$ of the QD, the linewidth $\Gamma$ of the eigenmodes and thereby their coupling $V \propto \sqrt{\Gamma}$ to the bulk superconducting electrons increases, which is a well-known effect due to increased surface-bulk scattering[28,38]. These effects will be used in the following to continuously pitch QD eigenmodes with different couplings $V$ through $E_F$ by accordingly tuning $L_x$.

**Emergence of in-gap states**

We now focus on the low-energy properties of the QDs in the region of the superconducting gap. d$I$/d$V$ spectroscopy of the QD presented in Fig. 1d shows clean superconductor-insulator-superconductor (SIS) tunneling without any in-gap states at spatial locations where no QD eigenmodes are present (gray curve in Fig. 2a and gray cross in Fig. 1e): sharp and prominent peaks appear at bias voltages corresponding to $e·V = \pm(\Delta_t + \Delta_s)$, indicating tunneling between the coherence peaks of tip and sample. Only a weaker thermal resonance peak at $e·V = \pm(\Delta_t - \Delta_s)$, i.e., at roughly zero bias, is observed (the bias range $|e·V| < \Delta_t$ is left out in Figs. 2a,c, see Supplementary Note 2 for more details). This confirms that the bulk gap of Ag(111) is fully developed for the given island thickness. In contrast, when measuring on a maximum of the QD eigenmode closest to $E_F$, we find a pair of sharp electronic states at particle-hole symmetric energies $\pm(\Delta_t + \varepsilon_\pm)$ within the gap (blue curve in Fig. 2a and blue cross in Fig. 1e). When mapping the spatial distribution of these states (Fig. 2b), we find that they closely resemble the shape of the expected QD eigenmode at $E \approx E_F$ as obtained from particle-in-a-box simulations (rightmost panel). Similar sharp, individual in-gap states are also observed in naturally occurring Ag(111) regions exhibiting strong 2D confinement on the sample (see Supplementary Note 5). To gain more insight into the nature of these in-gap states, we tune the QD's length $L_x$ and study the evolution of both the eigenmodes outside and inside the gap (Fig. 2c). As expected, the eigenmodes with quantum numbers [$n_x$, 1] outside the gap move in energy following the well-known $L_x^{-2}$ behavior (white dashed lines, see also Supplementary Note 4). Moreover, it can be seen that the peaks at $\pm(\Delta_t + \Delta_s)$ (white vertical dashed lines) remain at the same energy for all QD sizes, indicating that they stem from the proximitized Ag bulk states. Most notably, it can be observed that the in-gap states at varying energies $\pm(\Delta_t + \varepsilon_\pm)$ appear whenever a QD eigenmode energy $E_r$ approaches $E_F$. The absolute value for $\varepsilon_\pm$ is lowest when the QD's length $L_x$ is such that the $E_r$ would cross $E_F$ if extrapolated from outside the superconducting gap to the energetical region inside the gap (see dashed lines in Fig. 2c).

We evaluate this minimum value $\varepsilon_{min}$ for different eigenmodes of the QC and compare the results with their estimated energetic broadening $\Gamma$ at energies outside of the gap (see Supplementary Note 4 for details on the analysis). The energetic broadening is known to be predominantly related to the inverse lifetime of quasiparticles in the respective QD eigenmode for energies close to $E_F$. Furthermore, as noted above, $\Gamma$ of the eigenmodes close to $E_F$ decreases with increased QD size[28,38]. Indeed, this trend can be seen in Fig. 2d for the eigenmodes with increasing $n_x$, i.e., for wider QDs. As a main result of this work, there is a clear correlation between $\varepsilon_{min}$ and $\Gamma \propto V^2$: For increased couplings $\Gamma$ of a zero-energy QD eigenmode to the substrate superconductor, $\varepsilon_{min}$ is shifted from the Fermi energy towards the substrate's gap edge $\Delta_s$, i.e., the QD eigenmode's gap is gradually getting wider (see Fig. 2d). In the following, the origin of the in-gap states will be theoretically investigated and the link to the superconducting proximity effect will be substantiated.

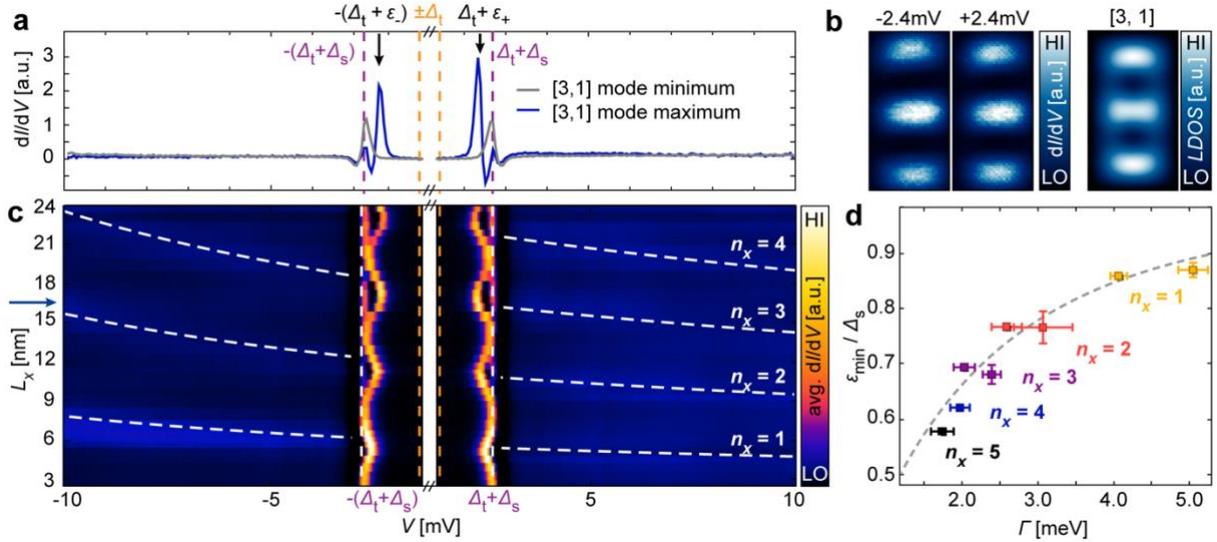

**Fig. 2 | In-gap states of near-zero-energy pitched QD eigenmodes. a**, d$I$/d$V$ spectra measured at two different positions in the QD with $L_x$ = 16.4 nm, $L_y$ = 9.1 nm shown in Figs. 1d-f. The respective positions are marked by gray and blue crosses in Fig. 1e. Both spectra were acquired at the same tip height. The values of the tip's superconducting gap $eV = \pm\Delta_t$ and the sum $eV = \pm(\Delta_t + \Delta_s)$ of the tip gap and the proximity induced Ag bulk gap $\Delta_s$ are marked by dashed orange and purple lines, respectively. The bias range $|e \cdot V| < \Delta_t$ is left out in a and c, see Supplementary Note 2 for examples of spectra over the full voltage range. In-gap states appear at particle-hole symmetric energies $\pm(\Delta_t + \varepsilon_\pm)$, which are marked by black arrows. **b**, Left: Constant-height d$I$/d$V$ maps measured at the energies of the in-gap state peaks in the same area as in Fig. 1e. Right: Particle-in-a-box simulation evaluated at zero energy with dominant contribution of the eigenmode with [$n_x$, $n_y$] = [3, 1]. **c**, Evolution of averaged d$I$/d$V$ spectra from d$I$/d$V$ line-profiles measured along the central vertical axis of different QDs (see, e.g., dashed orange vertical lines in Figs. 1c and d) as a function of the QD length $L_x$. The white dashed lines mark the evolution of the eigenmodes with $n_y = 1$ and $n_x = \{1, 2, 3, 4\}$ obtained from fitting the d$I$/d$V$ spectra at energies outside of the gap in Supplementary Note 4. The length of the QD presented in panels a and b is marked by the blue arrow on the left side. **d**, Linewidths $\Gamma$ of different QD eigenmodes extracted from fitting data from different QDs to Lorentzian peaks at energies outside of the gap. These are compared to the minimal energies of the in-gap states found when $E_r \approx 0$ (see Supplementary Note 4 for details of the fitting). The gray dashed line is the expected theoretical relation between $\varepsilon_{min}$ and $\Gamma$ for a spin-degenerate level coupled to a superconducting bath[12]. Parameters: $V_{stab}$ = -15 mV, $I_{stab}$ = 4 nA, $V_{mod}$ = 50 μV for panels a and c; $V_{stab}$ = -15 mV, $I_{stab}$ = 4 nA, $V_{mod}$ = 100 μV for panel b. Note, that further QDs constructed and analyzed as described in Supplementary Note 5 are included in panel d.

**Theoretical model of a spin-degenerate level coupled to a superconducting bath**

The observation of in-gap states is a surprising result, since particle-hole-symmetric states inside the gap of an *s*-wave superconductor are commonly believed to only appear for magnetic impurities[40,41]. In-gap states emerging around non-magnetic impurities are mostly considered to be evidence for unconventional SC[42,43]. In our samples, we exclude that magnetism plays a role on the pure and well-characterized noble-metal surface with only non-magnetic adatoms. Furthermore, Nb is a conventional *s*-wave superconductor and the proximity effect induced in a normal metal with negligible spin-orbit coupling is not expected to induce considerable unconventional pairing. However, as shown theoretically 50 years ago by Machida and Shibata[12], there is always a sub-gap solution for the problem of a localized spin-degenerate level, as present in the QDs in our samples, coupled to a superconducting bath due to resonance scattering[12,40]. We consider the Hamiltonian of Machida and Shibata in Ref. [12], i.e.,

$$\mathcal{H} = \sum_{k,\sigma} \epsilon_k c_{k,\sigma}^\dagger c_{k,\sigma} + \sum_{k,\sigma} V(c_{k,\sigma}^\dagger d_\sigma + d_\sigma^\dagger c_{k,\sigma}) + \sum_\sigma E_r d_\sigma^\dagger d_\sigma - \Delta_s \sum_k (c_{k,\uparrow}^\dagger c_{-k,\downarrow}^\dagger + c_{-k,\downarrow} c_{k,\uparrow}), \quad (1)$$

where $c_{k,\sigma}$ ($c_{k,\sigma}^\dagger$) and $d_\sigma$ ($d_\sigma^\dagger$) refer to the annihilation (creation) operators of superconducting bath electrons and the localized level, respectively. $\epsilon_k$ denotes the superconductor's normal electronic dispersion, $V \propto \sqrt{\Gamma}$ is the coupling strength of the localized level at energy $E_r$ to the bath and $\Delta_s$ is the order parameter of *s*-wave superconductivity in the bath. Calculating the local density of states (LDOS) of the localized level (see Methods) confirms that there is always a pair of bound states at in-gap energies[12] for all nonvanishing $V$. In the following, we will refer to these states as Machida-Shibata-States (MSSs). We depict the energy evolution of MSSs as a function of the localized level's energy $E_r$ in the normal state for different choices of $\Gamma$ in Fig. 3. For $\Gamma/\Delta_s \ll 1$ (Fig. 3a), the localized level couples only weakly to the superconductor and its energy $\varepsilon$ evolves mostly continuously through the gap while its particle-hole-symmetric partner state at -$\varepsilon$ features negligible spectral weight in the LDOS. As $\Gamma/\Delta_s$ is increased (Fig. 3b), the states with $\varepsilon_\pm$ show a pronounced anti-crossing behavior as $E_r$ approaches zero. Moreover, both states at $\varepsilon_\pm$ acquire a finite spectral weight in the LDOS, indicating that the superconductor mixes particle- and hole-like states. This situation is closely reminiscent of the experimental data in Fig. 2c. For strong coupling $\Gamma/\Delta_s \gg 1$, the in-gap states shift close to $\Delta_s$ irrespective of $E_r$, consistent with the regular proximity effect being induced into the localized resonance level leading to a full superconducting gap. We observe a similar effect in a tight-binding description of a QD weakly coupled to a superconducting surface layer (Supplementary Note 6), corroborating that the simplified description of the QD's eigenmode as a single localized quantum level $E_r$ shown in Fig. 3 is appropriate. The predicted shift of the MSSs' minimal energy with increasing $\Gamma$ is included as a grey dashed line in Fig. 2d. Its good quantitative agreement with the experimental data without additional fitting parameters suggests that the resonances found experimentally are indeed the previously unobserved MSSs.

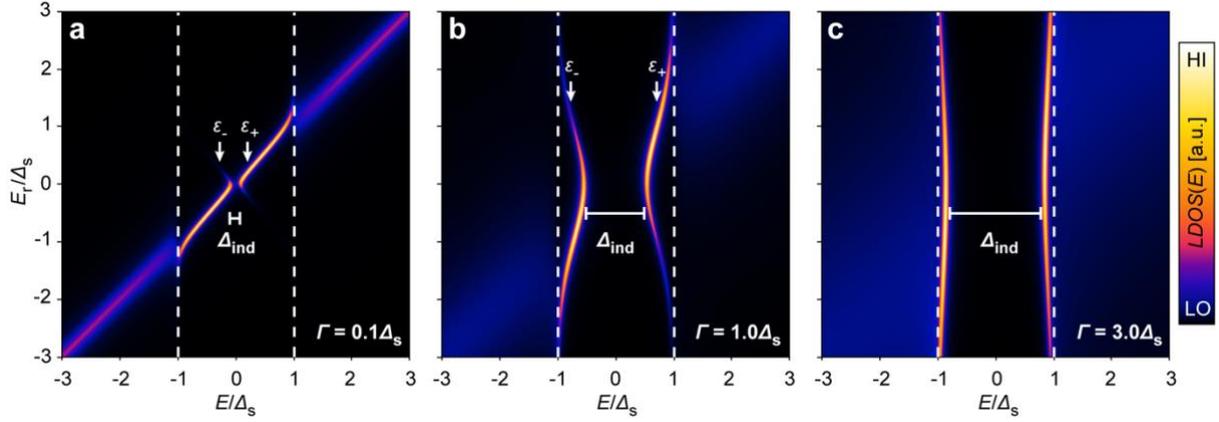

**Fig. 3 | Machida-Shibata states (MSS) from resonance scattering at a spin-degenerate localized level.**
**a**, Energy dependent local electron density of states LDOS($E$) of a single localized level at energy $E_r$ coupled to a superconducting bath with the parameter $\Delta_s$. The coupling strength $\Gamma \propto V^2$ (see Methods) equals $0.1\Delta_s$. The induced gap $\Delta_{ind}$ and the energies of the in-gap states $\varepsilon_\pm$ are marked. **b**, Same as panel a but for $\Gamma = 1.0\Delta_s$. **c**, Same as panel a but for $\Gamma = 3.0\Delta_s$. An energetic broadening of $\delta E = 0.03\Delta_s$ has been added in all panels (see Methods).

While these results demonstrate that the lowest-energy quasiparticle excitations of the local level become gradually gapped out with increasing coupling to the superconducting bath, it is not *a priori* clear whether the local level experiences proximity superconductivity. To this end, we perform a Schrieffer-Wolff transformation of Eq. (1) to obtain the effective low-energy theory of the level when $E_r$ lies within the superconductor's gap (see Supplementary Note 7 for details). The resulting Hamiltonian reads

$$\mathcal{H}'_D = \sum_\sigma E_r(1 + \Delta_{ind}/\Delta_s)d^\dagger_\sigma d_\sigma - (\Delta_{ind}d_\uparrow d_\downarrow + \text{h.c.}). \tag{2}$$

Indeed, it includes a term for the induced pairing energy $\Delta_{ind}$ of the level's quasiparticles, resembling the BCS-like mean-field-expression for superconductivity. Based on Eq. (2), it can be seen that, for $E_r = 0$, the lowest energy eigenstates of the system are energetically located at $\varepsilon_\pm = \pm\Delta_{ind}$. Thereby, the values of $\varepsilon_{min}$ we measured for the different QD eigenmodes (Fig. 2d) can indeed be identified with the proximity gap magnitudes $\Delta_{ind}$, which approach $\Delta_s$ for strong coupling $\Gamma$.

**Visualization of particle-hole-mixing**
Notably, the observed in-gap states at $\varepsilon_+$ and $\varepsilon_-$ are not symmetric in intensity. Their peak asymmetry in spectral weight can be analyzed in terms of the Bogoliubov mixing angle

$$\theta_B = \text{ArcTan}\left(\sqrt{|u|^2/|v|^2}\right) = \text{ArcTan}\left(\sqrt{A_{\varepsilon_+}/A_{\varepsilon_-}}\right). \tag{3}$$

Here, $u$ and $v$ are the respective particle- and hole-amplitudes of the Bogoliubov quasiparticles, which are related to the peak heights $A_{\varepsilon_\pm}$ at positive and negative peak energies $\varepsilon_\pm$ measured in tunneling spectroscopy[44]. The results are shown in Fig. 4. For maximal particle-hole mixing ($|u|^2 = |v|^2$), the angle $\theta_B$ equals $\pi/4$. For Bogoliubov quasiparticles, this case is expected when their energy approaches the pairing energy $\varepsilon_\pm \approx \pm\Delta_{ind}$. In the experimental data, we find a value of $\theta_B \approx \pi/4$ whenever $\bar{\varepsilon} \approx \varepsilon_{\min}$ ($E_r \approx 0$, see Supplementary Note 4). This finding further supports the above conjecture that $\varepsilon_{\min}$ can be interpreted as a proximity-induced superconducting pairing $\Delta_{\text{ind}}$ in the QD resonance level. When moving to larger in-gap state energies, $\theta_B$ either increases (for $E_r > 0$) or decreases (for $E_r < 0$). This trend is found consistently for all eigenmodes and qualitatively agrees well with the expectations for Bogoliubov excitation solutions of Eq. (2) (dashed gray line in Fig. 4).

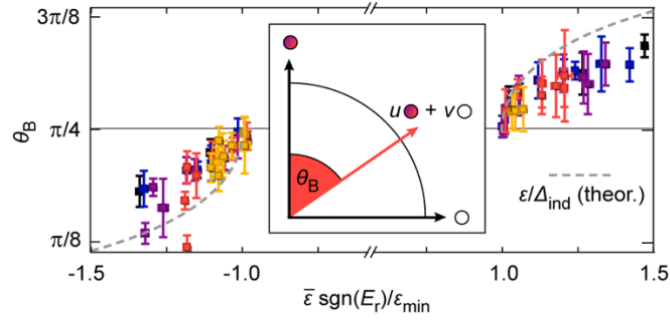

**Fig. 4 | Particle-hole mixture of the in-gap states.** Bogoliubov angle $\theta_B$ of the in-gap states with different energies $\bar{\varepsilon}$ normalized to their minimal energies $\varepsilon_{\min}$. The gray dashed line represents the expected relation for Bogoliubov quasiparticles with an induced gap of $\Delta_{\text{ind}} = \varepsilon_{\min}$ as derived from the effective Hamiltonian in Eq. (2) (see Supplementary Note 7 for details). Inset: Bogoliubov quasiparticles are coherent combinations of electrons (filled circle) and holes (empty circle). The Bogoliubov angle $\theta_B$ of a quasiparticle quantifies the amount of particle-hole mixing.

**Discussion**

The appearance of sharp in-gap states in STM experiments on superconductors is typically considered to be a fingerprint of either a local magnetic moment[40,41] or unconventional superconducting pairing[42,43]. Our experimental observation of MSSs clearly challenges this conclusion. The energy of the MSSs critically depends on the ratio of the resonance level linewidth $\Gamma$ measured outside of the superconducting gap and the superconducting pairing energy $\Delta_s$ present in the bath. For typical localized levels residing on single atomic impurities, this ratio is $\Gamma/\Delta_s \gg 1$ and thus the bound states are located at energies very close to the coherence peaks of the bath superconductor. Therefore, these resonances were previously neither studied extensively theoretically nor experimentally observed. However, the linewidths of the eigenmodes of the QDs studied here are of similar magnitude as the superconducting gap, which leads to the low-energy in-gap states depicted in Fig. 2c that are well split off from the coherence peaks. It should be noted that the spectral function displays two main features (see Fig. 3): the MSSs, which are – in theory – infinitely sharp in-gap states for all values of $\Gamma$ and the Lorentzian resonance level of linewidth $\Gamma$ outside of the gap. The measured linewidth of the MSSs is 80μeV, which is just at the border of our experimental resolution, so the intrinsic linewidth $\gamma$ of the MSS is even smaller, $\gamma < 80$μeV. In turn this implies that the lifetime of Bogoliubov quasiparticles in the MSSs exceeds $\hbar/\gamma = \tau > 8$ps at $T = 4.5$K. The sharpness of the in-gap states can be understood as a consequence of negligible scattering into the gapped bulk states, unlike in the metallic state at $E > \Delta_s$ where the level obtains a broadening $\Gamma \gg \gamma$.

As we have shown, the continuous shift of the MSSs to energies $\pm \Delta_s$ as the level's coupling to a superconductor is increased can be interpreted as the most miniature version of the proximity effect being induced into a single quantum level. The strongest coupling $\Gamma$ is observed for the narrowest investigated quantum dot ($n_x$ = 1 in Fig. 2d), resulting in a comparably large gap $\Delta_{\text{ind}}$ of up to 85% of $\Delta_s$ induced into the QD eigenmode. This strongly suggests that the proximity effect originates from scattering of the surface state at the QDs walls, which is maximal for the narrowest QDs, as also speculated by recent works[19,22]. Since the coupling $\Gamma$ is controlled by the QD size, the induced gap $\Delta_{\text{ind}}$ is found to be tunable as well (see Fig. 2d). Moreover, as demonstrated in Fig. 4, the experimentally observed resonance peaks behave like Bogoliubov excitations, which are expected to carry an energy dependent fractional charge[45] of $(|u|^2 - |v|^2)e$. This could potentially be directly probed by STM-based shot-noise measurements[46], opening avenues for studying quasiparticles with tunable fractional charge on the atomic-scale.

We anticipate that the concept of impurity-supported proximity-induced Cooper pairing in atom-by-atom designed quantum corrals could be helpful in general to induce SC into arbitrary surface states, potentially also combined with non-trivial topology. Amongst others, the latter presents a pathway for the creation of unconventional SC and Majorana bound states[5,6,19,22,47,48]. Moreover, patterning the surface states of (111) noble metal surfaces by precisely positioned scattering centers has evolved to one of the most promising platforms in the direction of artificial lattices. These have been shown to host Dirac fermions[49,50], flat bands[51–53], wavefunctions in fractal geometries[54] or topologically non-trivial states[53,55]. Eventually, our results facilitate studying the interaction of these exotic phenomena with superconducting pairing in a simple and tunable platform. Notably, while Coulomb interactions inside the noble metal QDs we study here are typically screened well by the charge carriers in the system's bulk, it would be interesting to extend this platform towards reduced screening, potentially enabling atomic scale studies of the crossover from spin-degenerate to spinful quantum dots coupled to superconductors[56].

**Methods**
**Experimental procedures**
The experiments were performed in a commercially purchased SPECS STM system operated at $T$ = 4.5 K which is equipped with home-built UHV chambers for sample preparation[57]. A Nb(110) single crystal was used as a substrate and cleaned by high temperature flashes to $T \approx$ 2000 K with an e-beam heater. As shown previously[36], this method yields an ordered but oxygen-reconstructed Nb(110) surface. Ag was deposited from an e-beam evaporator using a high-purity rod at a deposition rate of about 0.1 monolayers (ML)/min. In agreement with previous studies, evaporation of Ag at elevated temperatures leads to the formation of two pseudomorphic monolayers of Ag followed by Stranski-Krastanov growth of large Ag(111) islands (see Supplementary Note 1). In order to get preferably small and thin islands, we grew Ag islands in a three-step process, starting with the deposition of 2MLs at $T \approx$ 600 K creating two closed wetting layers. In a second step, the temperature was reduced to $T \approx$ 400 K to limit the lateral diffusion of Ag on the surface and to create more nucleation centers for the Stranski-Krastanov islands. Under these conditions, another 2MLs of Ag were deposited, followed by three additional MLs grown at $T \approx$ 600 K again to guarantee a well annealed surface of the topmost layers. STM images were obtained by regulating the tunneling current $I_{stab}$ to a constant value with a feedback loop while applying a constant bias voltage $V_{stab}$ across the tunneling junction. For measurements of differential tunneling conductance (d$I$/d$V$) spectra, the tip was stabilized at bias voltage $V_{stab}$ and current $I_{stab}$ as individually noted in the figure captions. In a next step, the feedback loop was switched off and the bias voltage was swept from -$V_{stab}$ to +$V_{stab}$. The d$I$/d$V$ signal was measured using standard lock-in techniques with a small modulation voltage $V_{mod}$ (RMS) of frequency $f$ = 1097.1Hz added to $V_{stab}$. d$I$/d$V$ line-profiles were acquired recording multiple d$I$/d$V$ spectra along a one-dimensional line of lateral positions on the sample, respectively. Note that the tip was not stabilized again after each individual spectrum was acquired but the line-profiles were measured in constant-height mode. This avoids artifacts stemming from a modulated stabilization height. At the chosen stabilization parameters, the contribution of multiple Andreev reflections and direct Cooper pair tunneling to the superconducting tip can be neglected (see Supplementary Note 2). Throughout this work, we use Nb tips made from a mechanically cut and sharpened high-purity Nb wire. The tips were flashed in situ to about 1500 K to remove residual contaminants or oxide layers. The use of superconducting tips increases the effective energy resolution of the experiment beyond the Fermi-Dirac limit[58] but requires careful interpretation of the acquired d$I$/d$V$ spectra. These are proportional to the convolution of the sample's LDOS, the superconducting tip DOS and the difference of the Fermi-Dirac distributions of tip and sample. Notably, the latter can play a large role when measuring at $T$ = 4.5 K. Details on the interpretation of SIS tunneling spectra and on the determination of the tip's superconducting gap $\Delta_t$ can be found in Supplementary Note 2. Ag atoms were reproducibly extracted out of the Ag(111) surface by approaching the tip to the surface as shown in Ref. [39] and Supplementary Note 3. Ag QDs were constructed by lateral atom manipulation[59,60] at low tunneling resistances of $R \approx 100$ kΩ.

**Theoretical model for resonance scattering at a spin-degenerate level coupled to a superconducting bath**

We consider a system of a single spin-degenerate local level coupled to an *s*-wave superconducting 3D bath, following the model introduced in Ref. [12] and the Hamiltonian given in Eq. (1). We calculate the LDOS at the local level. For that, we use the Green functions' equations of motion in energy space[61]

$$E G_{ij}(E) = \delta_{ij} + \langle\langle [o_i, \mathcal{H}]; o_j^\dagger \rangle\rangle, \quad (4)$$

where $o_i$ ($o_i^\dagger$) is an electron annihilation (creation) operator and $G_{ij}(E) = \langle\langle o_i; o_j^\dagger \rangle\rangle$ is the shorthand notation for the usual retarded Green's function[61]. The LDOS at the local level is

$$\text{LDOS}(E) = -\frac{1}{\pi} \text{Im}\left[\left(G_{d_\uparrow, d_\uparrow^\dagger}(\omega) + G_{d_\downarrow, d_\downarrow^\dagger}(\omega)\right)/2\right], \quad (5)$$

where $\omega = E + i \cdot \delta E$ and $\delta E$ is a small and positive real number. By solving the system of equations of motion Eq. (4) for the Hamiltonian in Eq. (1), we find

$$G_{d_\sigma, d_\sigma^\dagger}(\omega) = \frac{\omega + E_r + \dfrac{\Gamma \omega}{\sqrt{\Delta_s^2 - \omega^2}}}{\omega^2 \left(1 + \dfrac{2\Gamma}{\sqrt{\Delta_s^2 - \omega^2}}\right) - E_r^2 - \Gamma^2}, \quad (6)$$

where $\Gamma = \pi V^2 D$, $D = k_F^2 W / (2\pi^2 v_F)$ is the density of states per spin species of the substrate above the critical temperature at the Fermi level, and $W$ is its volume. Furthermore, $\sigma$ represents the spin up and down contribution, respectively, and the standard approximation of linearizing $\epsilon_k = v_F (k - k_F)$ around the Fermi energy has been used, with $v_F$ and $k_F$ being the Fermi velocity and momentum, respectively. The LDOS is given by

$$\text{LDOS}(E) = -\frac{1}{\pi} \text{Im}\left[\frac{\omega + E_r + \dfrac{\Gamma \omega}{\sqrt{\Delta_s^2 - \omega^2}}}{\omega^2 \left(1 + \dfrac{2\Gamma}{\sqrt{\Delta_s^2 - \omega^2}}\right) - E_r^2 - \Gamma^2}\right]. \quad (7)$$

We note the emergence of in-gap states as found in Ref. [12]. In contrast to a metallic bath, where the scattering results in a spectral broadening of the local level, the superconducting bath induces superconductivity by proximity to the local level. Hence, when $E_r$ lies within the superconductor's gap, the state at $E_r$ splits into two particle-hole symmetric ones around $E_F$. Notably, for energy scales much larger than $\Delta_s$, Eq.(7) reduces to a typical Lorentzian LDOS of width $\Gamma$ at position $E_r$, as observed in the experiment.

The obtained spin-degenerate single-level Hamiltonian with proximity induced pairing (Eq. (2)) is equivalent to the Green's function approach above to second order in the coupling constant $V \propto \sqrt{\Gamma}$. Assuming a spherical Fermi surface, we derive the explicit form of the induced superconducting term and the correction to the chemical potential of the quantum level in Supplementary Note 7.

**Data availability**
The authors declare that the data supporting the findings of this study are available within the paper and its supplementary information files.

**Code availability**
The analysis codes that support the findings of the study are available from the corresponding authors on reasonable request.

**Acknowledgements**
L.S., I.I., J.N-S., J.W., and R.W. gratefully acknowledge funding by the Cluster of Excellence 'Advanced Imaging of Matter' (EXC 2056 - project ID 390715994) of the Deutsche Forschungsgemeinschaft (DFG). K.T.T, J.W. and R.W. acknowledge support by the DFG – SFB-925 – project 170620586. T.P. acknowledges support by the DFG (project no. 420120155). R.W. acknowledges funding of the European Union via the ERC Advanced Grant ADMIRE (project No. 786020). We thank Philip Beck, Howon Kim, Roberto Lo Conte, Tim Wehling, Michael Potthoff and Alexander Weismann for helpful discussions.

**Author contributions**
L.S. and J.W. conceived the experiments. L.S. and K.T.T. performed the measurements and analyzed the experimental data. L.S. simulated the QD eigenmodes using the hard-wall model. I.I. derived the resonance scattering model and J.N.S. performed the numerical tight-binding simulations, both under the supervision of T.P.. T.P. derived the low-energy model. L.S. prepared the figures, L.S. and J.W. wrote the manuscript. All authors contributed to the discussions and to correcting the manuscript.

**Competing interests**
The authors declare no competing interests.